\documentclass[showpacs,aps,prc,preprint,nofootinbib,superscriptaddress]{revtex4}

\usepackage{epsfig}
\usepackage{amsfonts}
\usepackage{amsmath}
\usepackage{amssymb}
\usepackage{slashed}
\usepackage{color}
\usepackage{hhline}
\usepackage[addedmarkup=uwave,deletedmarkup=sout]{changes}
\newcommand{\bea}{\begin{eqnarray}}
\newcommand{\eea}{\end{eqnarray}}
\def\la{\langle}
\def\ra{\rangle}
\newcommand{\sll}[1]{#1\hspace{-0.5em}/}
\def\bm{\boldsymbol}
\def\vs{{\bm\sigma}}
\def\vr{{\bm r}}

\newcommand{\ext}{\textrm{ext}}

\newcommand{\ket}[1]{\lvert #1 \rangle}
\newcommand{\bra}[1]{\langle #1 \rvert}

\newcommand{\Psik}{\Psi_{1,\vec{k}} }

\begin{document}

\title{Screening of Nucleon Electric Dipole Moments in Nuclei}

\author{Satoru Inoue}
\email{inoue@mailbox.sc.edu}
\author{Vladimir Gudkov}
\email{gudkov@sc.edu}
\author{Matthias R.~Schindler}
\email{mschindl@mailbox.sc.edu}
\affiliation{Department of Physics and Astronomy,\\
University of South Carolina,\\
Columbia, SC 29208}
\author{Young-Ho Song}
\email{yhsong@ibs.re.kr}
\affiliation{Rare Isotope Science Project, Institute for Basic Science, Daejeon 305-811, Korea}
%\thanks{}

\date{\today}

\begin{abstract}

A partial screening of nucleon electric dipole moments (EDMs) in nuclear systems,
which is related to the Schiff mechanism known for neutral atomic systems, is discussed. 
It is shown that  the direct contribution from the neutron EDM
to the deuteron EDM is partially screened by about 1\% in a 
zero-range approximation calculation.

\end{abstract}

\pacs{24.80.+y,  11.30.Er,  21.10.Ky}

%\keywords{}

\maketitle

%%%%%%%%%%%%%%%%%%%%%%%%%%%%%%%%%%%%%%%%%%%%%%%%%%%%%%%%%%%
\section{Introduction}
\label{sec:intro}

The possible observation of permanent electric dipole moments (EDMs),
which violate both time-reversal invariance and parity~\cite{Landau:1957},
can be important evidence of physics beyond the Standard Model, and EDMs have been the subject of intense experimental and theoretical investigations
for more than 50 years (see, for example, recent reviews  \cite{Dmitriev:2003hs,Pospelov:2005pr,Engel:2013lsa,Chupp:2014gka} and references therein).

The Schiff theorem (see \cite{Schiff:1963}
and its extensions~\cite{Sandars:1967,Flambaum:1984fb,Liu:Schiff,Auerbach:2008,Flambaum:2011yk})
states that in a neutral system of non-relativistic point-like particles with Coulomb interactions,
the particles' intrinsic electric dipole moments are completely screened.
As a consequence, in a neutral atomic system
the nuclear EDM is screened and only the residual EDM of the nucleus after screening,
known as the nuclear Schiff moment, can be observed.
Though the measurement of the nuclear EDM without electrons would have less
uncertainty from a theoretical point of view,
the acceleration of the charged nucleus in an electric field
made this approach impractical and most EDM searches have focused
on neutral systems. However, with the recent advance in
using the fine-tuned momentum technique in storage rings,
direct measurements of nuclear EDMs will be feasible
in the near future~\cite{Khriplovich:1998zq,Farley:2003wt,Semertzidis:2009zz,Lehrach:2012eg}.

The value of the nuclear EDM can be defined
(see, for example~\cite{Liu:2004tq,Song:2012yh} and references therein)  as
\bea
\vec{d}=\la JJ|\hat{D}_{\sll{T}\sll{P}}|JJ\ra  ,
\eea
where $|JJ\ra$ is a nuclear state with total spin $J$ and its projection also equal to $J$.
The EDM operator $\hat{D}_{\sll{T}\sll{P}}$ contains direct contributions
from the intrinsic nucleon EDMs,
\bea
\label{eq:intr_edm}
\hat{D}^{nucleon}_{\sll{T}\sll{P}}
=\sum_i \frac{1}{2}[(d_p+d_n)+(d_p-d_n)\tau_i^z]\vs_i
\eea
and contributions from the nuclear EDM polarization operator,
\bea
\hat{D}^{pol}_{\sll{T}\sll{P}}
=\sum_i Q_i \vr_i ,
\eea
which describes the polarization of the nucleus due to time reversal invariance violating (TRIV) potentials.
Here, $d_n$ and $d_p$ are the neutron and proton EDMs,
and $Q_i$ and $r_i$ are the charge and position of the $i$-th nucleon in center-of-mass coordinates.
A non-zero nuclear EDM results in an energy shift of the system
in an external electric field.
Usually, contributions from the  polarization operator are larger then contributions
from the intrinsic nucleon EDMs, however both contributions are important.

In this paper we show that a partial screening of intrinsic EDMs, similar
to the one considered in the Schiff theorem, can occur in a charged system
of particles which also interact by strong interactions.
The origin of this effect lies in the interactions of the individual  EDMs
 with the electric field created by the charged particles in the system.
While there has been recent work  to
calculate the EDMs of light nuclei~\cite{Liu:2004tq,deVries:2011an,Song:2012yh,Bsaisou:2012rg,Bsaisou:2014zwa,Bsaisou:2014oka,Yamanaka:2015qfa},
 this effect has not been considered.
A one-photon exchange contribution to the nucleon-nucleon potential in which one of the vertices is the nucleon EDM was derived in an effective field theory framework in Ref.~\cite{Maekawa:2011vs}, but its effects have not been explicitly considered in subsequent calculations.
We will estimate the magnitude of this screening in the simple case of deuterons, using first a zero-range approximation, and then a square well potential. Finally, we discuss the generalization to the case of larger nuclei, where this screening effect also exists.

%%%%%%%%%%%%%%%%%%%%%%%%%%%%%%%%%%%%%%%%%%%%%%%%%%%%%%%%%%%
\section{Schiff Theorem}
\label{sec:theorem}
For completeness we present a short proof of the Schiff theorem
that applies to neutral systems, and can be adapted to charged systems (see~\cite{Flambaum:2011yk} for an example of the Schiff theorem applied to ions).
The total Hamiltonian of non-relativistic particles in a constant electric $\vec{E}$ field, interacting through electrostatic forces can be written as:
\begin{equation}\label{eq:Htot}
H = T + V_{C-C} + V_{C-D} + V_{C}^{\ext} + V_{D}^{\ext},
\end{equation}
where 
\begin{align}
T = & - \sum_{i=1}^{N} \frac{\vec{\nabla}_{i}^{2}}{2 m_{i}} ,\\
V_{C-C} = & \frac{1}{2} \sum_{i\neq j} \frac{Q_{i} Q_{j}}{| \vec{x}_{i} - \vec{x}_{j}|} , \\
V_{C-D} = & \sum_{i \neq j} Q_{i} \vec{d}_{j} \cdot \vec{\nabla}_{j} \frac{1}{| \vec{x}_{i} - \vec{x}_{j}|} = - \sum_{i \neq j} Q_{i} \vec{d}_{j} \cdot \vec{\nabla}_{i} \frac{1}{| \vec{x}_{i} - \vec{x}_{j}|} ,\label{VCD} \\
V_{C}^{\ext} = & - \sum_{i} Q_{i} \vec{x}_{i} \cdot \vec{E} \label{Vext}  , \\
V_{D}^{\ext} = & - \sum_{i} \vec{d}_{i} \cdot \vec{E},
\end{align}
and ${\vec d}_i$ is the intrinsic EDM of the $i$-th particle.

We consider the unperturbed
Hamiltonian (which is different from the Hamiltonian in  Ref.~\cite{Schiff:1963}, but analogous to the approach of Ref.~\cite{Liu:Schiff})
\begin{equation}
H_{0} = T + V_{C-C},
\end{equation}
assuming that the energy levels of the unperturbed system are known,
\begin{equation} \label{eq:H0}
H_{0} \ket{n} = E_{n} \ket{n}.
\end{equation}
To demonstrate the screening effect, we calculate the energy shift $\Delta E_{n}$ due to the potential $V = V_{C-D} + V_{C}^{\ext} + V_{D}^{\ext}$
that is linear in both the external field and the intrinsic EDMs,
\begin{equation}\label{Eq:deltaE}
\Delta E_{2} = \bra{n} V_{D}^{\ext} \ket{n} + \left( \bra{n} V_{C}^{\ext} \sum_{m \neq n} \frac{\ket{m} \bra{m}}{E_{n} - E_{m}} V_{C-D} \ket{n} + \bra{n} V_{C-D} \sum_{m \neq n} \frac{\ket{m} \bra{m}}{E_{n} - E_{m}} V_{C}^{\ext} \ket{n} \right).
\end{equation}
We refer to the first term on the right-hand side of Eq.~\eqref{Eq:deltaE} as the direct term, and to the remaining terms
as the indirect one.
Introducing the displacement operator (note that here all particles are charged)
\begin{equation}
A \equiv \sum_{i} \frac{\vec{d}_{i} \cdot \vec{\nabla}_{i}}{Q_{i}},
\end{equation}
which commutes with $T$ and satisfies the following operator identities,
\begin{align}
\label{eq:comm1}
[A, V_{C-C}] = & V_{C-D}, \\
\label{eq:comm2}
[A, V_{C}^{\ext}] = & V_{D}^{\ext},
\end{align}
one can re-write the indirect term  as
\begin{align}\label{Eq:deltaE2}
&\bra{n} V_{C}^{\ext} \sum_{m \neq n} \frac{\ket{m} \bra{m}}{E_{n} - E_{m}} V_{C-D} \ket{n} + \bra{n} V_{C-D} \sum_{m \neq n} \frac{\ket{m} \bra{m}}{E_{n} - E_{m}} V_{C}^{\ext} \ket{n} \nonumber \\
& =  \bra{n} V_{C}^{\ext} \sum_{m \neq n} \frac{\ket{m} \bra{m}}{E_{n} - E_{m}} [A, H_{0}] \ket{n} + \bra{n} [A, H_{0}] \sum_{m \neq n} \frac{\ket{m} \bra{m}}{E_{n} - E_{m}} V_{C}^{\ext} \ket{n} \nonumber \\
& =  \sum_{m \neq n} \left( \bra{n} V_{C}^{\ext} \ket{m} \bra{m} A \ket{n} - \bra{n} A \ket{m} \bra{m} V_{C}^{\ext} \ket{n} \right) \nonumber \\
& =  \bra{n} - [A, V_{C}^{\ext}] \ket{n} \nonumber \\
& = - \bra{n} V_{D}^{\ext} \ket{n}.
\end{align}
This term exactly cancels the direct contribution and thus proves Schiff's theorem.
Note that the potential $V_{C-D}$, which describes the interactions of intrinsic particle EDMs with the electric fields from other particles, is essential
for the proof of the cancellation. 

To estimate the possible cancellation of nucleon EDMs in nuclei we 
apply a similar formalism.
However, in this case one has to take into account some additional features: (a) the neutrality of some of the constituents (neutrons), (b) the acceleration
of the whole charged system (nucleus) in the external electric field; and (c) the presence of strong nucleon-nucleon interactions.
Unfortunately, strong interactions cannot be treated analytically by introducing an explicit strong interaction potential into Eqs. (\ref{eq:Htot}) and (\ref{eq:H0}). Therefore, the energy shift of Eq.~(\ref{Eq:deltaE}) has to be calculated numerically for each particular nucleus. 
In this paper we consider the deuteron, which is the simplest system proposed
for nuclear EDM measurements
in storage rings~\cite{Khriplovich:1998zq,Farley:2003wt,Semertzidis:2009zz,Lehrach:2012eg}.
In future work, we plan to further investigate the screening of EDMs in heavier nuclei,
which would be required for precision calculations of nuclear EDMs.

%%%%%%%%%%%%%%%%%%%%%%%%%%%%%%%%%%%%%%%%%%%%%%%%%%%%%%%%%%%
\section{Deuteron}
\label{sec:deuteron}

The deuteron Hamiltonian  in an external constant electric field can be written as
\begin{equation}
H = \frac{p_{p}^{2}}{2m_{p}} + \frac{p_{n}^{2}}{2m_{n}} + V( \vec{x}_{p}, \vec{\sigma}_{p}, \vec{x}_{n}, \vec{\sigma}_{n}) + e \vec{d}_{n} \cdot \vec{\nabla}_{n}\left( \frac{1}{|\vec{x}_{p} - \vec{x}_{n}|}\right) - e \vec{x}_{p} \cdot \vec{E} - (\vec{d}_{p} + \vec{d}_{n}) \cdot \vec{E},
\end{equation}
where  $\vec{d}_p$ and $\vec{d}_n$ are the proton and neutron EDMs.
$V( \vec{x}_{p}, \vec{\sigma}_{p}, \vec{x}_{n}, \vec{\sigma}_{n})$ is the strong two-body potential that binds the nucleons.
Here we do not consider TRIV nucleon-nucleon interactions
because their contribution to the screening effect is of higher order in perturbation theory, and we find screening already without these interactions.
After separating the motion of the center-of-mass $\vec{X}$ by changing coordinate variables,
\begin{equation}
\vec{X} \equiv \frac{m_{p} \vec{x}_{p} + m_{n} \vec{x}_{n}}{m_{p} + m_{n}} , \ \ \vec{r} \equiv \vec{x}_{p} - \vec{x}_{n},
\end{equation}
we obtain the Hamiltonian
\begin{equation}
H = \frac{p_{X}^{2}}{2M} + \frac{p_{r}^{2}}{2\mu} + V(r, \vec{\sigma}_{p}, \vec{\sigma}_{n}) - e \vec{d}_{n} \cdot \vec{\nabla}_{r} \left( \frac{1}{r} \right) - e \vec{X} \cdot \vec{E} - \frac{em_{n}}{M} \vec{r} \cdot \vec{E} - (\vec{d}_{p} + \vec{d}_{n}) \cdot \vec{E},
\end{equation}
where
$M = m_{p} + m_{n}=2m_N$ is the total mass
and $\mu = m_{p} m_{n} / (m_{p} + m_{n})\approx m_N/2$ is the reduced mass,
while $m_N$ denotes the average nucleon mass. The center-of-mass motion is described by
\begin{equation}
H_{X} = \frac{p_{X}^{2}}{2M} - e \vec{X} \cdot \vec{E},
\end{equation}
and the remainder describes the relevant physics,
\begin{equation}
H_{r} = \frac{p_{r}^{2}}{2 \mu} + V(r, \vec{\sigma}_{p}, \vec{\sigma}_{n}) - e \vec{d}_{n} \cdot \vec{\nabla}_{r} \left( \frac{1}{r} \right) - \frac{em_{n}}{M} \vec{r} \cdot \vec{E} - (\vec{d}_{p} + \vec{d}_{n}) \cdot \vec{E}.
\end{equation}

To estimate the magnitude of the screening we consider the deuteron in the zero-range approximation. This model was first applied to deuteron EDM calculations in Ref.~\cite{Khriplovich:1999qr}. In this simplistic model the deuteron is described by the S-wave component of its wave function, which is taken as
\begin{equation}
\label{eq:ZRADeut}
\begin{split}
\psi_d(\vec{r}) & = R_0 (r) Y_0^0(\hat{r}) \\
& = \sqrt{\frac{\kappa}{2\pi}}\frac{e^{-\kappa r}}{r}\ .
\end{split}
\end{equation}
Here, $\kappa = \sqrt{m_N E_B}$ is the deuteron binding momentum, with  the deuteron binding energy $E_B=2.23\,\text{MeV}$. Since $V_{C}^{\ext}$ and $V_{C-D}^{\ext}$ carry orbital angular momentum of 1, intermediate states in Eq.~(\ref{Eq:deltaE}) must be $L=1$ components of scattering states. The zero-range potential does not affect $L=1$ states, and we can use the $L=1$ components of plane waves, which are
\begin{equation}
\begin{split}
\Psik(\vec{r}) & = R_{1,k} (r) \sum_{m=-1}^1 Y_1^{m*}(\hat{k}) Y_1^m(\hat{r}) \\
& = \frac{4\pi i}{(2\pi)^{3/2}}\, j_1(k r) \sum_{m=-1}^1 Y_1^{m*}(\hat{k}) Y_1^m(\hat{r}) \ .
\end{split}
\end{equation}
Thus, using these wave functions, the matrix elements $\langle 0 | \vec{r} \,| 1,\vec{k} \rangle$ and $\langle 0 | \frac{\vec{r}}{r^3} | 1,\vec{k} \rangle$ contributing to the second-order contribution in Eq.~\eqref{Eq:deltaE} can be calculated analytically.

The angular parts of the coordinate space integrals can be performed using the orthogonality of the spherical harmonics. The radial parts of the resulting matrix elements are given by
\begin{equation}
\label{eq:In}
I_n(k) = \int dr \,r^n \frac{e^{-\kappa r}}{r} \, j_1(kr) ,
\end{equation}
with $n=3$ and $n=0$, respectively,
resulting in
\begin{align}
I_3 (k) & =  \frac{2k}{(k^2+\kappa^2)^2} , \\
I_0 (k) & =  - \frac{\kappa}{2k} + \frac{k^2 + \kappa^2}{2k^2}\arctan\left( \frac{k}{\kappa} \right) .
\end{align}
Using these expressions as well as
\begin{equation}
E_0 = -E_B = -\frac{\kappa^2}{m_N} \ , \quad E_{1,k} = \frac{k^2}{m_N} \ ,
\end{equation}
the second-order energy shift $\Delta E_2$ is given by\footnote{We thank J.~de Vries, C.~Hanhart, A.~Nogga, and A.~Wirzba for pointing out a numerical error in our calculation.}
\begin{eqnarray}
\Delta E_2 &=&  \frac{4 e^{2} \kappa m_N}{3 \pi} {\vec d}_n\cdot{\vec E} \int_0^\infty dk \frac{k^2}{\kappa^2+k^2} I_{3}(k) I_{0}(k) \notag \\
& = &  \frac{\alpha}{12} \sqrt{\frac{m_N}{E_B}} \, \vec{d_n}\cdot \vec{E} \notag \\
& \approx & 0.013 \, \vec{d_n}\cdot \vec{E} \ .\label{eq:DeutRes}
\end{eqnarray}
Therefore, in the deuteron the direct contribution from the neutron EDM is partially screened by about 1\%. This is consistent with the power counting estimate that can be made from Ref.~\cite{Maekawa:2011vs}, in which TV potentials based on one-photon exchange as well as on one-pion exchange are derived. Comparing the spin-isospin structure of the two potentials and using the result for the one-pion-exchange contribution to the deuteron EDM derived in Ref.~\cite{deVries:2011re}, one can estimate the effect considered here to be about 1\%.
Because the neutron is not charged, there is no corresponding screening of the proton EDM.
As seen from Eq.~\eqref{eq:DeutRes}, the value of the  screening would be reduced if the deuteron were more deeply bound.
Strictly speaking, however, the zero-range approximation rests on the fact that the deuteron binding energy is small, and we cannot extrapolate to a deeply bound system. The power counting estimate would also not be appropriate for heavier systems in which the typical nucleon momenta are too large for the application of the EFT framework of Ref.~\cite{Maekawa:2011vs}.

To estimate the uncertainty of these calculations,  we include effective range effects in the deuteron wave function of Eq.~\eqref{eq:ZRADeut} by multiplying it by a factor of $(1-\kappa \rho_t)^{-1/2}$, where $\rho_t \approx 1.76\text{ fm}$ is the $^3S_1$ effective range. The second-order energy shift  increases to
\begin{equation}
\label{eq:DeutResMod}
\Delta E_2 \approx 0.021 \, \vec{d_n}\cdot \vec{E} \ ,
\end{equation}
which is comparable with the zero-range approximation result of Eq.~(\ref{eq:DeutRes}).

To study how deviations from the zero-range potential can affect the size of the  screening, we consider a square well potential as a model of the deuteron,
\begin{equation}
V(\vec{r}) = \left\{ \begin{array}{ll} -V_{0}, & r<r_{0} \\
                 0, & r \geq r_{0} \ , \end{array} \right.
\end{equation}
 with the parameters tuned to reproduce the deuteron binding energy $E_B = 2.23 \, \text{MeV}$.
In this model the deuteron radial wave function  is
\begin{equation}
R_{0}(r) = \left\{ \begin{array}{ll} N_{1} j_{0}(k' r) \ , & r<r_{0} \\
                                     N_{2} \frac{e^{-\kappa r}}{\kappa r}, & r>r_{0} \ ,
                                     \end{array} \right.
\end{equation}
where
\begin{equation}
E_{0} = -E_{B} = -\frac{\kappa^{2}}{m_N} \ , \quad V_{0} - E_{0} = \frac{k'^{2}}{m_N} \ .
\end{equation}
Continuity of $R_0(r)$ and its derivative at $r=r_{0}$ requires the relation
\begin{equation}
\label{eq:FixV0}
k' \cot k'r_{0} = -\kappa \ .
\end{equation}
For a given $r_{0}$, we fix $V_{0}$ to be the smallest positive energy that satisfies Eq.~\eqref{eq:FixV0}.\footnote{There are infinitely many solutions to $V_{0}$ in this equation, but other solutions correspond to deeper potentials with an excited state with binding energy $E_B$.} The normalization coefficients  $N_{1,2}$ are also fixed numerically so that the wave function is continuous and normalized to 1.

The $L=1$ component of the scattering wave function is given by
\begin{equation}
\begin{split}
\Psik(\vec{r}) & = R_{1,k} (r) \sum_{m=-1}^1 Y_1^{m*}(\hat{k}) Y_1^m(\hat{r})\\
& = \frac{4\pi i}{(2\pi)^{3/2}}\, \sum_{m=-1}^1 Y_1^{m*}(\hat{k}) Y_1^m(\hat{r}) \left\{
\begin{array}{ll} N_{3} j_{1}(k'' r) , & r<r_{0} \\
                  \left[ j_{1}(kr) + ikf_{1}(k) h_{1}^{(1)}(kr) \right], & r>r_{0},
                  \end{array} \right.  ,
\end{split}
\end{equation}
where $k'' = \sqrt{m_N (V_{0} + E)}$. The scattering amplitude, $f_1$, and the normalization constant, $N_3$, are again found from continuity conditions. This solution is exact for a square well.

To determine the energy shift, we calculate the matrix elements $I_3$ and $I_0$ as defined in Eq.~\eqref{eq:In}.
The numerical results for the second-order energy shift  $\Delta E_2$ for different values of $r_0$ are given in Table~\ref{table:DeltaE}.
As expected, the result converges to the zero-range value as $r_0 \rightarrow 0$.
Despite the considerable $r_0$ dependence of $\Delta E_2$ in the square well model, all values are within the expected range given by the estimates of Eqs.~\eqref{eq:DeutRes} and \eqref{eq:DeutResMod}, and the screening is least in the simplistic zero-range approximation. Therefore, there is no indication that this effect  can vanish in an improved treatment of deuteron and P-wave scattering wave functions.

\begin{table}
\begin{center}
\caption{The second-order energy shift, $\Delta E_2$, for selected values of $r_0$. The second column is the analytic zero-range result.} \label{table:DeltaE}
\begin{tabular}{|c||c|c|c|c|c|c|}
\hline
$r_0$ (fm) & 0 & 0.1 & 0.5 & 1.0 & 1.5 & 2.0 \\
\hline
$\Delta E_2$ ($\vec{d}_n \cdot \vec{E}$) & 0.013 & 0.013 & 0.014 & 0.015 & 0.017 & 0.018 \\
\hline
\end{tabular}
\end{center}
\end{table}

%%%%%%%%%%%%%%%%%%%%%%%%%%%%%%%%%%%%%%%%%%%%%%%%%%%%%

\section{Discussion}

The above results show the existence of a Schiff-type screening of the neutron EDM contribution to the deuteron EDM. In the zero-range approximation, this contribution is reduced by roughly 1\%.
The size of the effect is at least in part due to the fact that the deuteron binding energy is very small.
Within this approximation, the second-order energy shift $\Delta E_2$ is proportional to $E_B^{-1/2}$; an increase in the binding energy may reduce the screening. However, the applicability of the zero-range approximation relies on the smallness of the binding energy.

For heavier nuclei, one must also consider the screening of proton EDM contributions due to electric fields from other protons.
Our Eq.~(\ref{Eq:deltaE}) applies for any nucleus, and the terms in parentheses give the size of the screening effect in general.
$V_{C-D}$ contains both proton and neutron EDMs, and in general they can both be screened; for the deuteron the proton EDM is not screened because there are no other charged particles.
In the case of heavier nuclei, the strong NN potential cannot be modeled in a simple form and solved analytically.
For an accurate answer, a numerical treatment for a given nucleus is required.
However, we do find a scaling argument for the size of screening for heavier nuclei. The isovector nature of $V_{C}^{\ext}$ and $V_{C-D}$ suggests that the intermediate states in Eq.~(\ref{Eq:deltaE}) would be dominated by the giant dipole resonance~\cite{Goldhaber:1948zza, Bohr}. We considered a toy model for the dipole resonance, where proton and neutron distributions are each treated as uniform spheres, and the relative coordinate between the centers of the distributions undergoes harmonic oscillator motion. Scalings for the potentials in Eq.~(\ref{Eq:deltaE}), the resonance frequency, and the nuclear radius are all known, and we find the second-order energy shift $\Delta E_2$ to scale as $A^{2/3}$. This suggests that the screening effect would grow for larger nuclei, and that it should be considered in future calculations of nuclear EDMs.

\begin{acknowledgments}
We thank J.~de Vries, C.~Hanhart, A.~Nogga, and A.~Wirzba for  pointing out the power-counting argument to estimate the expected size of the effect considered here.
This material is based upon work supported by the U.S. Department of Energy Office of Science,
Office of Nuclear Physics program under Award Number DE-FG02-09ER41621 (VG)
and Award Number DE-SC0010300 (SI and MRS).
The work of YS was supported by the Rare Isotope Science Project of the Institute for Basic Science
funded by the Ministry of Science, ICT and the
Future Planning and National Research Foundation of Korea (2013M7A1A1075764).
\end{acknowledgments}

\bibliography{TViolation}

\end{document}